\documentclass{article}
\usepackage{amsmath,amssymb,amsfonts,spconf}
\usepackage{enumitem}
\setlist{nosep, leftmargin=14pt}

\usepackage{cite}
\usepackage{pifont}
\usepackage{algorithmic}
\usepackage{graphicx}
\usepackage{textcomp}
\usepackage{xcolor}
\usepackage{xspace}
\usepackage[colorlinks=true,linkcolor=blue,citecolor=blue, urlcolor=blue]{hyperref}

\def\BibTeX{{\rm B\kern-.05em{\sc i\kern-.025em b}\kern-.08em
    T\kern-.1667em\lower.7ex\hbox{E}\kern-.125emX}}
\usepackage{tabularray}

\usepackage[normalem]{ulem} 

\definecolor{commentred}{RGB}{240,0,0} 

\def\sd#1{\color{black} {\scriptsize $\pm$ \hspace{-1.4mm} #1} \color{black}}

\begin{document}

\title{ConUNETR: A conditional transformer Network for 3D Micro-CT Embryonic Cartilage Segmentation}

\name{Nishchal Sapkota$^{\dagger}$, \hspace{-4mm}
\quad 
Yejia Zhang$^{\dagger}$, \hspace{-4mm}
\quad 
Susan M. Motch Perrine$^{\star}$,  \hspace{-3mm}
\quad 
Yuhan Hsi$^{\star}$, 
\quad Sirui Li$^{\dagger}$,
\quad Meng Wu$^{\S\#}$, \vspace{-0.48cm}
}
\address{
\textit{Greg Holmes}$^{\#}$, \hspace{-4.9mm}
\quad 
\textit{Abdul R. Abdulai}$^{\#}$,  \hspace{-4.2mm}
\quad 
\textit{Ethylin W. Jabs}$^{\S\#}$,  \hspace{-4.2mm}
\quad 
\textit{Joan T. Richtsmeier}$^{\star}$,  \hspace{-4.9mm}
\quad 
\textit{Danny Z. Chen}$^{\dagger}$ \\ \\
$^{\dagger}$Department of Computer Science and Engineering, University of Notre Dame, IN 46556 \\
$^\star$Department of Anthropology, The Pennsylvania State University, PA 16801 \\ 
$^\S$Department of Clinical Genomics, Mayo Clinic, Rochester, MN 55905 \\
$^{\#}$Department of Genetics and Genomic Sciences, Icahn School of Medicine at Mount Sinai, NY 10029
}

\maketitle
\begin{abstract} 
Studying the morphological development of cartilaginous and osseous structures is critical to the early detection of life-threatening skeletal dysmorphology. Embryonic cartilage undergoes rapid structural changes within hours, introducing biological variations and morphological shifts that limit the generalization of deep learning-based segmentation models that infer across multiple embryonic age groups. Obtaining individual models for each age group is expensive and less effective, while direct transfer (predicting an age unseen during training) suffers a potential performance drop due to morphological shifts. We propose a novel Transformer-based segmentation model with improved biological priors that better distills morphologically diverse information through conditional mechanisms. This enables a single model to accurately predict cartilage across multiple age groups. Experiments on the mice cartilage dataset show the superiority of our new model compared to other competitive segmentation models. Additional studies on a separate mice cartilage dataset with a distinct mutation show that our model generalizes well and effectively captures age-based cartilage morphology patterns.
\end{abstract}


\section{Introduction}

Congenital anomalies (birth defects) that include cartilaginous and osseous (bone) structures are a major cause of infant mortality and childhood morbidity, affecting 2-3\% of human neonates \cite{mossey2003global}. 
Accurate segmentation of cartilaginous structures (e.g., the chondrocranium and Meckel's cartilage ~\cite{pitirri2022meckel, motch2023embryonic}) is central to understanding these diseases and revealing novel targets for treatment.
Embryonic mice are often sacrificed at different prenatal periods due to the invasive nature of these studies, to study cartilage development through contrast-enhanced micro-CT. 

During embryonic growth,
large and rapid changes in cartilage size and morphology occur, making the automatic segmentation of cartilage challenging.
More specifically, many \textit{morphological variations} such as discontinuous or absent cartilage structures, poorly identifiable boundaries, and different cartilage thickening or thinning are observed during growth. 
Moreover, studying multi-aged mice with different mutations (the usual practice in this line of work) adds more challenges because of the various morphological differences among even same-age cartilages across mutations.
Due to the difficulty of encapsulating cartilage diversity across age groups and mutations, separate models are often trained for each age group and each mutation.
However, this approach requires intensive annotation labor, incurs higher training costs/efforts, and risks overfitting to a single age group/mutation (see Figure \ref{moreDataPlotSameAge}) without leveraging inter-age/mutation structure information.


Alternatively, training a single model on the combined set of all age groups offers the benefits of an expanded training set and better generalizability.
This idea has been explored with proven results in biomedical image analysis with ``universal models'' \cite{universalUnet,universalLesion} that attempt to train a single model to perform well across modalities, biological structures, or image acquisition settings. 
However, this approach remains unexplored in cross-sectional studies like our setting where the goal is to model structural changes over time rather than over different modalities and structures of interest.
Further, enabling our model to generalize to unseen age groups and other mutation cohorts is an additional challenge that would be valuable for craniofacial researchers. 


\begin{figure*}[h!] 
\begin{center} 
\includegraphics[width=0.79\linewidth]{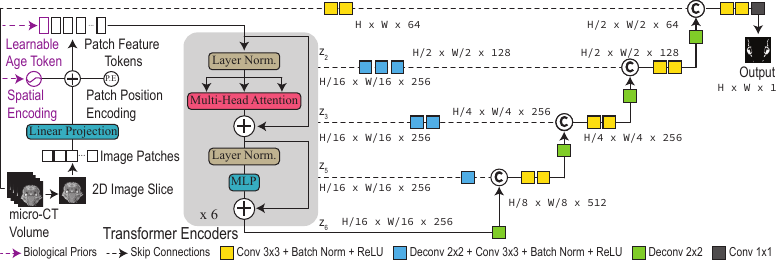} 
\caption{\small{An overview of the architecture of our proposed Conditional Universal Model (ConUNETR).}}
\label{main diagram} 
\end{center}
\end{figure*} 

Related to embryonic cartilage analysis, knee or hip cartilage segmentation are well-studied problems \cite{poistionPriorClusterKneeCartSeg, liu2022isegformer,shapePriorsCart, li2023sdmt, peng2022kcb, lin2022calibrating, zeng2020entropyHip}.
While these models have successfully addressed multi-domain adaptation and proposed novel ways to add cartilage shape priors to facilitate segmentation even in very small data regimes, the structure of interests are well-developed in their cases and do not contain longitudinal samples with drastic morphology variations like in our developing embryonic cartilages cases.
Embryonic mouse cartilage segmentation has seen fewer studies, and known methods either use iterative active learning with experts in the loop \cite{motch2023embryonic} or account for morphological variations only in the skull regions of the same age mice  \cite{hao2020cartilage}. 
To this end, our work is the first to address morphology variations across multi-age and multi-mutation embryonic cartilage data with a single ``universal" model.

In this paper, we propose a new cartilage segmentation model, ConUNETR, to enable accurate prediction across age groups and improve generalizability to unseen ages and mutations. 
We leverage attention in Transformers~\cite {transformer} to effectively capture global image context while informing feature extraction components of relevant biological priors. 
We propose to add these priors as conditional inputs to the Transformer encoder (i.e., ViT~\cite{vit}) in the form of tokens so that the network can extract age-relevant features during the forward pass and model cartilage structure variations across time for improved generalizability. Besides, we propose spatial embedding components to tackle the lack of spatial context from non-consecutive 2D labeled slices of micro-CT volumes.

Our \textbf{main contributions} are: (1) a novel conditional modeling approach using biological priors in Transformer encoders to address morphology variations in the multi-age embryonic cartilage dataset;  
(2) a lightweight 2D Transformer segmentation network with spatial awareness for non-consecutive 2D labeled slices of micro-CT volumes;
(3) superior generalizability across age groups and across mutations compared to popular segmentation models.
\begin{table*}[h]
\begin{center}
\caption{\label{ind} \textbf{Individual Age-group Training and Direct Transfer}: In Tables 1 and 2, the numbers represent the per volume dice scores averaged within an age group (average across 3 runs). Grey columns represent the cases with the model trained on age-specific training data, and white columns represent the ones with the model transferred directly with no training data.}
\resizebox{0.80\textwidth}{!}{
\begin{tblr}{    
    columns={colsep=3pt},
    colspec={l|c c c| c c c| c c c },
}
\hline
\hline
\multirow{2}{15mm}{Model} & \multicolumn{3}{c} {Trained on MUTA-13.5} & \multicolumn{3}{c} {Trained on MUTA-14.5} & \multicolumn{3}{c} {Trained on MUTA-15.5}\\
\cline{2-11} 
& \SetCell[c=1]{bg=gray!20!black!10}MUTA-13.5 & MUTA-14.5 & MUTA-15.5 & MUTA-13.5 & \SetCell[c=1]{bg=gray!20!black!10}MUTA-14.5 & MUTA-15.5 & MUTA-13.5 & MUTA-14.5 & \SetCell[c=1]{bg=gray!20!black!10}MUTA-15.5 \\
\hline
\hline
U-Net 
& \SetCell[c=1]{bg=gray!20!black!10} \textbf{\underline{71.1}}  \sd{1.3}
& 63.1  \sd{1.4}
& 39.7  \sd{2.2}

& 47.9  \sd{4.0}
& \SetCell[c=1]{bg=gray!20!black!10} 89.6  \sd{0.7}
& 71.7  \sd{4.9}

& 32.5  \sd{2.1}
& 79.0  \sd{1.3}
& \SetCell[c=1]{bg=gray!20!black!10}83.0 \sd{0.8}\\ 

Res2Unet 
& \SetCell[c=1]{bg=gray!20!black!10}69.5   \sd{0.9}
& 62.0   \sd{1.8}
& 26.2   \sd{4.0}

& 46.3   \sd{2.6}
& \SetCell[c=1]{bg=gray!20!black!10} 89.2   \sd{0.9}
& 68.3   \sd{1.4}

& 33.5   \sd{2.3}
& 81.3   \sd{0.7}
& \SetCell[c=1]{bg=gray!20!black!10} \textbf{\underline{83.7}} \sd{0.9} \\ 
UNETR
& \SetCell[c=1]{bg=gray!20!black!10}68.2   \sd{0.2}
& 61.0   \sd{1.9}
& 36.1   \sd{2.3}

& 36.1   \sd{5.2}
& \SetCell[c=1]{bg=gray!20!black!10}\textbf{\underline{90.1}}   \sd{0.5} 
& 63.7   \sd{1.7}

&31.3   \sd{2.6}
& 78.1   \sd{0.3}
& \SetCell[c=1]{bg=gray!20!black!10} 82.2 \sd{0.5}\\
\hline
\end{tblr}
}
\caption{\label{joint} \textbf{Joint Age-group Training and Cross Mutation Generability}}
\resizebox{1.0\textwidth}{!}{
\begin{tblr}{
    columns={colsep=2pt},
    colspec={l|c c c| c c c| c c c | c c },
}
\hline
\hline
\multirow{2}{15mm}{Model} & \multicolumn{3}{c} {Trained on MUTA-13.5 \& 14.5} & \multicolumn{3}{c} {Trained on MUTA-14.5 \& 15.5} & \multicolumn{3}{c} {Trained on All MUTA}  & \multicolumn{2}{c} {Trained on All MUTA}\\
\cline{2-12} 
& \SetCell[c=1]{bg=gray!20!black!10}MUTA-13.5 & \SetCell[c=1]{bg=gray!20!black!10}MUTA-14.5 & MUTA-15.5 & MUTA-13.5 & \SetCell[c=1]{bg=gray!20!black!10}MUTA-14.5 & \SetCell[c=1]{bg=gray!20!black!10}MUTA-15.5 & \SetCell[c=1]{bg=gray!20!black!10}MUTA-13.5 & \SetCell[c=1]{bg=gray!20!black!10}MUTA-14.5 & \SetCell[c=1]{bg=gray!20!black!10}MUTA-15.5 & ACH-14.5 & ACH-16.5 \\
\hline
\hline
U-Net 
& \SetCell[c=1]{bg=gray!20!black!10}70.7 \sd{0.5}
& \SetCell[c=1]{bg=gray!20!black!10}89.4  \sd{0.9}
& \textbf{\underline{70.5}}  \sd{1.9}

& 46.8  \sd{2.0}
& \SetCell[c=1]{bg=gray!20!black!10}89.4  \sd{0.6}
& \SetCell[c=1]{bg=gray!20!black!10}84.7  \sd{0.8}

& \SetCell[c=1]{bg=gray!20!black!10}71.0  \sd{1.3}
& \SetCell[c=1]{bg=gray!20!black!10}89.8 \sd{0.4}
& \SetCell[c=1]{bg=gray!20!black!10}85.0 \sd{1.2}

& 67.1 \sd{2.3}
& 65.6 \sd{2.5} \\ 
Res2Unet 
& \SetCell[c=1]{bg=gray!20!black!10}69.9  \sd{1.3}
& \SetCell[c=1]{bg=gray!20!black!10} 89.5  \sd{0.8}
& 67.8  \sd{1.2}

& 45.2  \sd{0.7}
& \SetCell[c=1]{bg=gray!20!black!10}89.7  \sd{0.6}
& \SetCell[c=1]{bg=gray!20!black!10}\textbf{\underline{85.0}}  \sd{0.3}

& \SetCell[c=1]{bg=gray!20!black!10}72.0  \sd{0.4}
& \SetCell[c=1]{bg=gray!20!black!10}90.3  \sd{0.8}
& \SetCell[c=1]{bg=gray!20!black!10}84.1 \sd{0.3}

& 79.4 \sd{2.2}
& 78.8  \sd{1.0} \\ 
UNETR
& \SetCell[c=1]{bg=gray!20!black!10}69.9  \sd{0.6}
& \SetCell[c=1]{bg=gray!20!black!10}90.1   \sd{0.4}
& 64.6  \sd{1.3}

& 47.2  \sd{1.9}
& \SetCell[c=1]{bg=gray!20!black!10}90.5  \sd{0.8}
& \SetCell[c=1]{bg=gray!20!black!10}84.0  \sd{1.2}

& \SetCell[c=1]{bg=gray!20!black!10}70.2   \sd{0.2}
& \SetCell[c=1]{bg=gray!20!black!10}90.6  \sd{0.5}
& \SetCell[c=1]{bg=gray!20!black!10}83.6 \sd{0.6}

& 78.1 \sd{1.2}
& 75.1 \sd{3.3} \\
\hline
ours
& \SetCell[c=1]{bg=gray!20!black!10} \textbf{\underline{72.5}} \sd{0.6}  
& \SetCell[c=1]{bg=gray!20!black!10} \textbf{\underline{90.9}} \sd{0.3}  
& 70.0 \sd{0.5} 

& \textbf{\underline{48.1}} \sd{2.3}
&\SetCell[c=1]{bg=gray!20!black!10} \textbf{\underline{91.4}} \sd{0.3}
& \SetCell[c=1]{bg=gray!20!black!10} 84.6  \sd{1.0}

& \SetCell[c=1]{bg=gray!20!black!10}\textbf{\underline{73.3}} \sd{0.8}  
&\SetCell[c=1]{bg=gray!20!black!10} \textbf{\underline{91.7}} \sd{0.3}
& \SetCell[c=1]{bg=gray!20!black!10}\textbf{\underline{85.4}}  \sd{0.2}

& \textbf{\underline{82.3}} \sd{1.1}
& \textbf{\underline{84.0}}  \sd{1.6} \\
\hline
\end{tblr}
}
\end{center}
\end{table*}
\vspace{-4mm}
\section{Methodology}
\vspace{-2mm}
We aim to build a universal cartilage segmentation model for multi-age embryonic data using biological priors and conditioning the model to learn age-specific morphology features to benefit from joint training on the multi-age data. Our segmentation model ConUNETR is inspired by the widely used UNETR~\cite{hatamizadeh2022unetr}, and through careful design changes, we achieve a lightweight network (12.3M parameters vs.~UNETR's 92.6M) to adapt for 
small training data regimes. Our model consists of Transformer encoders, CNN decoders, learnable \textit{age tokens}, and spatial encoding modules as shown in Figure~\ref{main diagram}. These components cost a few additional parameters and are seamlessly integrated with the Transformer encoder.

\subsection{Architecture}
Our ConUNETR model consists of stacked ViT-like \cite{vit} Transformer encoders with residual connections and CNN decoders with U-Net-like \cite{UNET} skip connections. 

\textbf{Encoder: }The encoder includes 6 stacked Transformers (6 stages), with 4 attention heads each, and maintains hidden dimensions ($d_{model}$) of 256. 
Given a 2D image of size $H\times W \times C$ from a volume, we divide it into non-overlapping patches of size $P\times P$ to obtain $N=HW/P^2$ patches. We flatten these patches to attain a 1D vector that is linearly projected to a $K$-dimensional embedding space and obtain patch embeddings ($E^1, \ldots, E^N$). Learnable patch position encoding ($pos_{i}$) for each $i^{th}$ patch, and sinusoidal relative slice spatial encoding ($e_{sp}$) (explained in Section~\ref{sec:spatial_encoding}) are added to the patch embeddings to obtain:
\vspace{-1mm}
$$e = [E^1;E^2;\ldots;E^N] + [pos_{1};pos_{2};\ldots;pos_{N}] + e_{sp}.$$
Age token $E^a$ (explained in Section~\ref{sec:age_tokens}) in the embedding space is prepended to the embedding vector $e$ to obtain:
$$z_0 = [E^a;\widetilde{E^1};\widetilde{E^2};\ldots;\widetilde{E^N}],$$ where $\widetilde{E^i}$ represents the $i^{th}$ patch embedding with position and spatial encoding added to it.
The first stage of the stacked Transformer encoder is fed with $z_0$ and it outputs $z_1$ after multi-headed self-attention is applied among features of embedded patch tokens and age tokens. Outputs of stages 2, 3, 5, and 6 ($z_2, z_3, z_5, z_6$) are passed to the decoder via skip connections after removing the age token feature ($z_i[0]$) and only image features are fed to the decoder. These image features have the morphology priors instilled in them due to self-attention with age token features.

\textbf{Decoder:}
The bottleneck layer at the end of the encoder upsamples the reshaped encoder output ($z_6$) by a factor of 2 using a deconvolution layer.
Multi-scale features $z_2, z_3, z_5$ are reshaped and converted from the embedding to the input space using convolutions and merged with the respective decoder blocks through skip connections. The rest of the decoder consists of 4 stages of upsampling blocks that receive feature maps from the encoder which are concatenated with the upsampled features. Softmax is applied to the output of the final stage of the decoder to obtain the segmentation maps.

\subsection{Age Tokens}
\label{sec:age_tokens}
We introduce an additional learnable token (similar in implementation to a \textit{class} token in \cite{vit}) to represent the age of samples in the dataset. $k$ age tokens are initialized to represent 
$k$ ages and the one corresponding to the age of the input image is prepended to the list of embedded patch features. 
We also try injecting the biological priors of the input image throughout the network using learnable age embeddings. Different from the age tokens which make use of biological priors in the encoder only, age embeddings allow us to represent concept information throughout the network by encoding this into the image patches directly (through pixel-wise summation). We have observed that using biological priors with tokens works better than embedding.

\subsection{Spatial Encoding}
\label{sec:spatial_encoding}
To overcome the loss of spatial contexts when operating on some 2D slices of a volume (due to non-consecutive slice annotations), we embed spatial encoding into linearly projected image patches. Note that, while patch position encoding represents the relative location of a patch within a 2D slice, Spatial Encoding represents the relative location of a 2D slice within a volume. We map these 2D slice locations from 1 to 100, and for each location ($loc$), build sinusoid wavelengths, forming a geometric progression from $2 \pi$ to $10000* 2\pi$.
The corresponding spatial encoding ($e_{sp}$) is added to each patch embedding. 
We also tried learnable spatial encoding but observed that sinusoid spatial encoding works slightly better.
\vspace{-1.5mm}
\begin{equation*}
\label{spatial-embd}
\begin{aligned}
\textit{Spatial Encoding}_{(loc, 2i)} &= sin(loc/10000^{2i/d_{model}}),\\
\textit{Spatial Encoding}_{(loc, 2i+1)} &= cos(loc/10000^{2i/d_{model}}).
\end{aligned}
\end{equation*}

\vspace{-3mm}
\section{Datasets and Implementation Details}
\vspace{-2mm}
\textbf{Dataset}: 
Mouse embryos were stained with phosphotungstic acid (PTA), and whole body micro-CT volumes ($\sim$1600 2D slices per volume) for mice of embryonic (E) ages E13.5, E14.5, and E15.5 days with craniofacial mutation A (MUTA) and ages E14.5 and E16.5 for 
\textit{Fgfr3$^{Y367C/+}$} mouse model with
achondroplasia (ACH) mutation \cite{motch2023embryonic} were acquired.
The training set for our experiments consists of 9 volumes (3 volumes per MUTA age) and the training ground truth consists of cartilage masks obtained through an iterative active learning framework \cite{motch2023embryonic} that produces cartilage predictions as good as experts' hand segmentations. Unless mentioned otherwise, we use only 5\% of image slices per volume (a common annotation budget in practice for our embryonic cartilage segmentation) during training. The test set includes the experts' hand segmentations on 2D slices and comprises 10 partially labeled ($\sim$5\%) volumes (2 volumes for each age in both mutations). Thus, our total training data consists of 
720 MUTA and the test set consists of 
480 MUTA and 
320 ACH 2D slices. We have made the ACH dataset
\footnote{\scriptsize{www.datacommons.psu.edu/commonswizard/MetadataDisplay.aspx?Dataset=6367}} 
publicly available in \cite{motch2023embryonic}.

\textbf{Implementation Details}:
All models are trained from scratch using the AdamW optimizer (initial learning rate of 0.0001) with weight decay of 0.001 and Cosine Annealing learning rate scheduler. Cross-entropy loss is applied to all the models, trained on a single NVIDIA A10 GPU for 700 epochs (batch size of 45) using AMP \cite{mixedPreTrain}. 512x512 sized image crops are obtained with augmentations including random cropping, a non-linear Bézier curve intensity transformation, horizontal/vertical flips, and $90^{\circ}$ rotations.

\textbf{Baselines:
} We compare the performance of our segmentation model (ViT encoder with spatial encoding and age tokens) with three popular models: (1) A small U-Net \cite{UNET} (7.9M parameters) with CNN encoder; (2) Res2UNet (25.4M) with a Res2Net \cite{gao2019res2net} encoder; (3) Small UNETR (12.3M) \cite{hatamizadeh2022unetr} with Transformer \cite{vit} encoder. A U-Net-like decoder with skip connections is added to all these encoders.

\vspace{-2mm}
\section{Experiments, Studies, and Results}

\subsection{Study on Same Age Group Training}

\begin{figure}[h!] 
\begin{center} 
\includegraphics[width=1.0\linewidth]{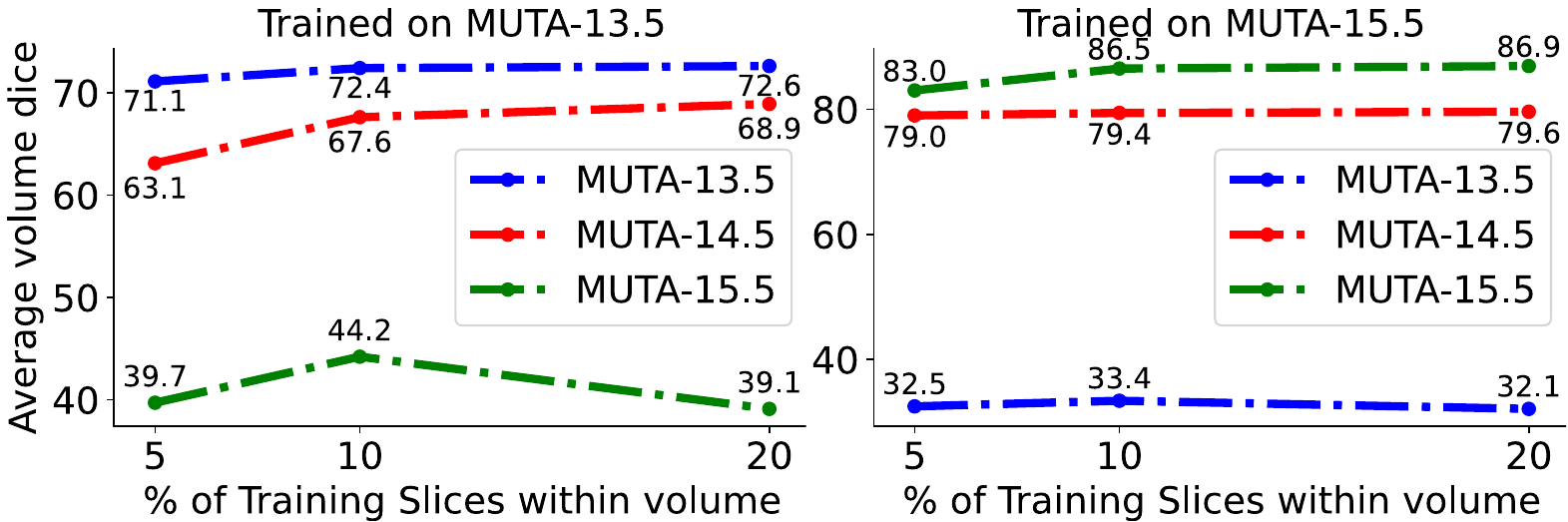} 
\vspace{-6mm}
\caption{\small{Performance with additional same-age train.~data (U-Net). 
}}
\label{moreDataPlotSameAge} 
\vspace{-4mm}
\end{center}
\end{figure}

Table \ref{ind} shows that the model trained on one age group does not generalize well to other age groups without additional training data, due to age-specific structural differences of developing cartilages. One might associate this to the small size of the training data but our experiments (see Figure \ref{moreDataPlotSameAge}) suggest that with more training data, a model trained on an individual age overfits on that age and does not generalize well across age groups. This emphasizes the need for joint training.
\vspace{-2mm}
\vspace{-5mm}
\subsection{Study on Joint Training}
The first two blocks in Table \ref{joint} show results with joint training on volumes with closer age groups, and the third block shows joint training on all age groups within the same mutation (MUTA). Except in some cases, all the models exhibit performance gains with joint training.
Cases with lower or subpar performance improvement from individual to joint training are explained by the difficulty in learning multi-age morphology variations. However, our model generally outperforms the baselines in two age-group joint training and beats all the baselines in three age-group training, indicating its stronger ability to utilize cross-age-group information.
\vspace{-4mm}
\subsection{Study on Cross Mutation Generalizability}
\vspace{-2mm}
We introduce additional mutation variations using ACH volumes. Models trained on all the three MUTA ages are applied without any ACH training data to ACH volumes. As seen in the $4^{th}$ block of Table \ref{joint}, U-Net yields poor generalizability while Res2Unet and UNETR yield subpar performance. Our conditional model, ConUNETR,  exhibits significant improvement over these baselines (2.9\% on E14.5 age and 5.2\% on E16.5 age volumes). Cartilages aged E16.5 are well developed and easily distinguishable from background structures, which explains the high performance on these volumes. 
Nevertheless, these improvements show that our model generalizes well in cross-mutation morphology variations.
\begin{figure}[h!] 
\begin{center} 
\includegraphics[width=0.61\linewidth]{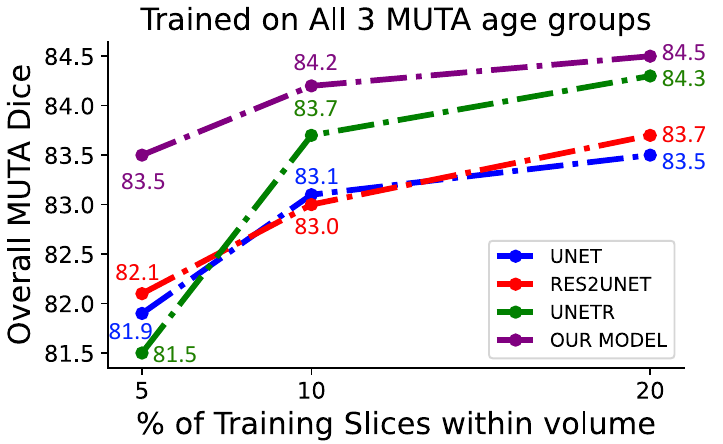} 
\vspace{-3mm}
\caption{\label{moreTraningData}
 \small{Performance with additional joint cross-age train. data (all MUTA ages). Average Dice score computed across all age groups.
}}
\end{center}
\end{figure} 
\vspace{-3mm}
\subsection{Study on Performance with Increased Training Data}
In Figure \ref{moreTraningData}, we observe that the CNN model has a lower performance ceiling while Transformer models make steady gains with additional joint age training data. The performance differences between the Transformers and our conditional Transformer model narrow as the training data size increases, suggesting that large training data potentially negate the need for biological priors. However, in practice, since the cartilage segmentation task is usually in a small data regime, our proposed model finds a meaningful spot in this domain. 
\vspace{-3mm}
\subsection{Ablation Study}
 Adding spatial embeddings to the ViT encoder yields 3.1\% on ACH and 0.7\% performance gain on MUTA (Table \ref{ablation}). The two conditional components, age embeddings and age tokens, contribute to an additional 2.7\% and 3.5\% respectively on multi-age ACH, and 0.5\% and 1.3\% respectively on multi-age MUTA volumes. These components add negligible differences in the total parameter count and training time.
\vspace{-3mm}
\begin{table}[h]
\begin{center}
\caption{\label{ablation} \small{\textbf{Ablation Study}: Average Dice scores across MUTA and ACH age groups obtained by the model trained on all MUTA ages.}}
\resizebox{0.80\linewidth}{!}{
\begin{tblr}{columns={colsep=3pt},
    colspec={l l l l | c | c  },
}
\multirow{2}{*}{\rotatebox{10}{ViT Encoder}} & 
\multirow{2}{*}{\rotatebox{10}{Spatial Encoding.}} & 
\multirow{2}{*}{\rotatebox{10}{Age Emb.}} & 
\multirow{2}{*}{\rotatebox{10}{Age Token}} & 
\multicolumn{2}{c}\\
& & &  & MUTA & ACH \\
\hline  
 \checkmark &&&&81.5 & 76.6  \\
   \checkmark & \checkmark &   &&82.2 & 79.7 \\
    \hline
  \checkmark &\checkmark& \checkmark && 82.7 & 82.4   \\
  \checkmark &\checkmark&  & \checkmark& 83.5 & 83.2   \\  
\hline
\end{tblr}
}
\end{center}
\end{table}

\vspace{-6mm}
\section{Conclusions}
\vspace{-3mm}
We studied the potential to use a universal model in segmenting multi-age/mutation embryonic mouse cartilages through conditional mechanisms using morphology and spatial priors. Our proposed lightweight segmentation model uses components that are seamlessly integrated with the Transformer encoder with very few additional parameters. Through multiple studies, we showed that our method demonstrates superior generalizability on new ages/mutations (even in lesser data regimes) compared to the popular segmentation models. We plan to extend this work further in facilitating accurate cartilage segmentations in weakly supervised settings and possibly on other datasets with similar challenges.


\section{COMPLIANCE WITH ETHICAL STANDARDS}

Mice were produced, processed, and sacrificed in compliance with animal welfare guidelines approved by the Icahn School
of Medicine at Mount Sinai and the Pennsylvania State University to create the dataset

\section{Acknowledgments}
This work was supported in part by NIH/NICDR grants R01 DE027677 to JTR and DZC, R01 DE031439 to SMMP and EWJ, and R01 DE029832 to EWJ and SMMP.

\bibliographystyle{plain}
\bibliography{refs}{}

\end{document}